\documentclass[sts,preprint]{imsart}
\usepackage[utf8]{inputenc}

\usepackage{amsmath}
\usepackage{amsfonts}
\usepackage{color}
\usepackage{hyperref}
\hypersetup{
    unicode=false,          
    pdftoolbar=true,        
    pdfmenubar=true,        
    pdffitwindow=false,     
    pdfstartview={FitH},    
    colorlinks=true,       
    linkcolor=black,          
    citecolor=blue,        
    filecolor=magenta,      
    urlcolor=blue           
}
\usepackage{graphicx}
\usepackage{natbib}
\usepackage[mathlines, pagewise]{lineno}

\def\R{\mathbb R}

\def\E{{\rm E}}
\def\Var{{\rm var}}
\def\Cov{{\rm cov}}

\def\xb{{\rm x}}

\def\xb{{\bf x}}

\def\gammag{\boldsymbol{\gamma}}
\def\lambdag{\boldsymbol{\lambda}}

\def\Sigmag{\boldsymbol{\Sigma}}
\def\betag{\boldsymbol{\beta}}
\def\gammag{\boldsymbol{\gamma}}

\begin{document}
\begin{frontmatter}

\title{Probability Sampling Designs:  Principles for Choice of Design and Balancing}
\runtitle{Principles for Choice of Design}

\begin{aug}
  \author{\fnms{Yves}  \snm{Till\'e}\corref{}\ead[label=e1]{yves.tille@unine.ch}},
  \author{\fnms{Matthieu} \snm{Wilhelm}\ead[label=e2]{matthieu.wilhelm@unine.ch}}
  \runauthor{Y. Till\'e and M. Wilhelm}
  \affiliation{University of Neuch\^{a}tel}
  \address{Institut de Statistique, Université de Neuchâtel, Av. de Bellevaux 51, 2000 Neuchâtel, Switzerland,\newline
          \printead{e1,e2}}
\end{aug}

%
%
%
%
%
%
%

\begin{abstract}
The aim of this paper is twofold. First, three theoretical principles are formalized: randomization, overrepresentation and restriction. We develop these principles and give a rationale for their use in choosing the sampling design in a systematic way.
In the model-assisted framework, knowledge of the population is formalized by modelling the population and the sampling design is chosen accordingly. We show how the principles of overrepresentation and of restriction naturally arise from the modelling of the population. The balanced sampling then appears as a consequence of the modelling. Second, a review of probability balanced sampling is presented through the model-assisted framework. For some basic models, balanced sampling can be shown to be an optimal sampling design.  Emphasis is placed on new spatial sampling methods and their related models. An illustrative example shows the advantages of the different methods. Throughout the paper, various examples illustrate how the three principles can be applied in order to improve inference.

\end{abstract}


\end{frontmatter}

\section{Introduction}
Very early in the history of statistics, it appeared that censuses were unachievable in many practical situations. Thus the idea of using a subset of the target population to infer certain characteristics of the entire population naturally appeared. This idea can be traced back at least to Pierre-Simon Laplace \citep{lap:1847}. In the first half of the XX$^{\text{th}}$ century, it became clear that only random sampling can provide an unbiased estimate. \citet{kru:mos:80} provide a concise review of the history of probability sampling.

Classical references for sampling designs include \citet{suk:54}, \citet{coc:77}, \citet{jes:78}, and \citet{bre:han:83}, who gave a list of 50 methods to select a sample with unequal inclusion probabilities. More modern textbooks include \citet{sar:swe:wre:92}, \citet{til:06}, \citet{lohr2009sampling} and \citet{tho:12}. The more recent developments in survey sampling have been mainly motivated by new applications and new types of data, as for instance functional data \citep{car:jos:11}.

For \citet{haj:59}, a survey is always characterized by a strategy composed of a sampling design and of an estimator of the parameter of interest. In the present paper, we focus on the choice of sampling design while we restrict attention to the Narain-Horvitz-Thompson (NHT) estimator of the total \citep{nar:51,hor:tho:52}.
In the case of a design-based approach, apart from the estimator, the practitioner can choose how the sample is selected, i.e. she/he has to determine a sampling design. This is the core of the theory of design-based survey sampling. This choice is driven by both theoretical and practical aspects.

In this paper, three important principles are introduced: randomization, overrepresentation and restriction. The relevance of these principles is justified. We are probably not the first to highlight that those principles are desirable, but we would like to introduce and discuss them in a comprehensive and systematic way.

The randomization principle states that the sampling designs must be as random as possible. Indeed, the more random the sample is, the better the asymptotic approximations are \citep{ber:98d,ber:98c}. Since most of the quantification of  uncertainty is carried out using asymptotic results, this is a very important aspect. Another point is that very random sampling designs (which will be clarified later) are more robust \citep{gra:10}. The principle of overrepresentation suggests to preferentially select  units where the dispersion is larger.
The principle of restriction excludes very particular samples such as samples with empty categories or samples where the NHT-estimators of some auxiliary variables are far from the population total. In this way, samples that are either non-practical or known to be inaccurate are avoided. The restrictions could consist of only choosing fixed size samples for example.

When auxiliary information is available, it is desirable to include it in the sampling design in order to increase the precision of the estimates. In the design-based approach, the auxiliary information should be used when choosing the sampling design. A balanced sample is such that the estimated totals of the auxiliary variables are approximately equal to the true totals. Intuitively, this can be seen as an a priori calibration \citep{dev:sar:92}. The cube method \citep{dev:til:04a} is a way to implement a probability sampling design which is balanced with equal or unequal first-order inclusion probabilities. The cube method is then a direct implementation of the principles of overrepresentation and restriction since it enables us to select samples with given inclusion probabilities and at the same time balanced on totals of auxiliary variables.
Special emphasis is also placed on balanced sampling with spatial applications.

The suggested principles cannot be the only foundation for the choice of sampling design. Many other aspects are important such as the simplicity of the procedure, the quality of the data frame or a low rate of non-response. Thus, these general principles are not always applicable because of practical constraints. However, we recommend adopting an approach where general principles should be considered in order to improve the quality of a survey.

There is no intention to be exhaustive in the enumeration of all the recent advances that have contributed to survey sampling. Our intention is more to highlight that taking into account the aforementioned principles can be a motivation for both theoretical and practical advances and that this is well illustrated by balanced sampling.

The paper is organized as follows. In Section~\ref{sec:notation}, definitions and the notation are given. In Section~\ref{sec:basic}, the most basic sampling designs are briefly described. In Section~\ref{sec:principles}, some principles of sampling are proposed.
Section~\ref{sec::balance} describes balanced sampling and briefly present  the cube method.
 In Section~\ref{sec:model-assis}, we propose a model-assisted selection of sampling designs in light of those principles. In Section~\ref{sec::spat-samp}, we present new methods for spatial sampling. An illustrative example presented in Section~\ref{sec:ill-examp} enables us to compare these methods.
Finally, a discussion concludes the paper in Section~\ref{sec:disc}.

\section{Probability sampling and estimation}
\label{sec:notation}
In the following, a list sampling frame is supposed to be available.
Consider a population $U$ composed of $N$ units that are denoted by their order numbers so it can be written $U=\{1,\dots,k,\dots,N\}$. Let us denote by $\mathcal{S}$ the set of subsets of $U$, which has cardinality $2^N$. A sample without replacement is simply an element $s \in \mathcal{S}$, that is a subset of the population. Note that the empty set is a possible sample. A sampling design $p(.)$ is a probability distribution on $\mathcal{S}$
$$
p(s) \ge 0\mbox{ and }\sum_{s\in \mathcal{S}}p(s)=1.
$$
A random sample $S$ is obtained by selecting a sample $s$ with probability $p(s)$. Thus $\Pr(S=s)=p(s),$ for all $s\in \mathcal{S}$. Hence, $S$ denotes the random variable and $s$ the realization of it. The set $\{s\in \mathcal{S} : p(s)>0 \} \subset\mathcal{S}$ is called the support of the sampling design. For instance, one can consider $ \mathcal{S}_n=\{s\in \mathcal{S}|\#s=n\}$ for a sampling design of fixed sample size $n$.

The first-order inclusion probability $\pi_k$ is the probability of selecting the unit $k$. The joint inclusion probability $\pi_{k\ell}$ is the probability that two different units $k,\ell$ are selected together in the sample. They can be derived from the sampling design:
$$
\pi_k = \sum_{s\ni k}p(s) \mbox{ and }
\pi_{k\ell} = \sum_{s\supset \{k,\ell\}} p(s).
$$

The aim is to estimate a total
$$
Y =\sum_{k\in U}y_k
$$
of the values $y_k$ taken by the variable of interest on all the units of the population.

The total $Y$ can be estimated by the Narain-Horvitz-Thompson estimator \citep{nar:51,hor:tho:52}
$$
\widehat{Y}=\sum_{k\in S}\frac{y_k}{\pi_k}.
$$
If $\pi_k>0$ for all $k\in U$, this estimator is unbiased, i.e. $\E_p(\widehat{Y})=Y,$ where $\E_p(.)$ is the expectation under the sampling design $p(.)$.

Define
$$
\Delta_{k\ell}=
\left\{
\begin{array}{ll}
\pi_{k\ell}-\pi_{k}\pi_{\ell} & \mbox{ if }k\neq \ell \\
\pi_{k}(1-\pi_{k}) & \mbox{ if }k= \ell.
\end{array}
\right.
$$
The design variance $\Var_p(.)$ of the NHT-estimator is equal to:
$$
\Var_p\left(\widehat{Y}\right) = \sum_{k\in U}\sum_{\ell \in U} \frac{y_{k} y_{\ell}}{\pi_k\pi_{\ell}}\Delta_{k\ell}.
$$
When the sample size is fixed, this variance simplifies to
$$
\Var_p\left(\widehat{Y}\right) = - \frac{1}{2}\sum_{k\in U}\sum_{\substack{\ell\in U \\ k \neq \ell}}
\left(\frac{y_{k}}{\pi_{k}} - \frac{y_{\ell}}{\pi_{\ell}} \right)^2 \Delta_{k\ell}.
$$
Estimators can be derived from these two expressions. For the general case,
$$
\widehat{\Var}\left(\widehat{Y}\right) = \sum_{k\in S}\sum_{\ell \in S} \frac{y_{k} y_{\ell}}{\pi_{k}\pi_{\ell}}\frac{\Delta_{k\ell}}{\pi_{k\ell}},
$$
where $\pi_{kk}=\pi_{k}$. When the sample size is fixed, the variance estimator \citep{sen:53,yat:gru:53} is given by:
$$
\widehat{\Var}\left(\widehat{Y}\right) = -\frac{1}{2} \sum_{k\in S}\sum_{\substack{\ell\in S\\ k \neq \ell}}
\left(\frac{y_{k}}{\pi_{k}} - \frac{y_{\ell}}{\pi_{\ell}} \right)^2 \frac{\Delta_{k\ell}}{\pi_{k\ell}}.
$$
These estimators are both unbiased provided that $\pi_{k\ell}>0$, $k\neq \ell \in U$.

Provided that the first-order inclusion probabilities are positive, the NHT estimator is unbiased and the variance and the mean squared error are equal. Provided that the first and the second order inclusion probabilities are positive, the variance estimators give an unbiased estimation of the mean-squared error. It is usual to assume a normal distribution to quantify the uncertainty. In many sampling designs, the normality assumption is asymptotically valid. The rate of convergence depends on the entropy \citep{ber:98d,ber:98c}, which is roughly speaking, a measure of randomness. We further discuss the concept of entropy in Section~\ref{sec:entropy}.

\section{Some basic designs}
\label{sec:basic}
In the following, a list sampling frame is supposed to be available. In some situations, this may be not the case, as for instance in spatial sampling where the sampling frame can be a geographical region and the units a subdivision of this region.
The sampling designs presented in this Section are all implemented in various R packages \citep{R}. \citet[][chap. 3.7]{valliant2013practical} provide a review of the current R and SAS packages for survey sampling.

\subsection{Bernoulli sampling design}

In Bernoulli sampling, the units are independently selected according to independent Bernoulli random variables with the same inclusion probabilities
$\pi$. Then,
$$
p(s) =\pi^{n_s} (1-\pi)^{N-n_s} \ , \mbox{ for all }s \in \mathcal{S},
$$
where $n_s$ is the sample size of sample $s.$ The sample size is random and has a binomial distribution, i.e. $n_s\sim Bin(N,\pi).$ The sample size expectation is $N\pi$. The first-order inclusion probability is $\pi_k=\pi$ and the second order inclusion probability is equal to $\pi_{k\ell}=\pi^2$ for $k\neq \ell.$

\subsection{Poisson sampling design}

When the inclusion probabilities $\pi_k$ are unequal, the sampling design obtained by selecting the units with independent Bernoulli random variables with parameter $\pi_k$ is called Poisson sampling.
The sampling design is
$$
p(s) =\prod_{k\in s}\pi_k \prod_{k\notin s}(1-\pi_k) \ , \mbox{ for all }s \in \mathcal{S}.
$$
The inclusion probabilities are $\pi_k$ and $\pi_{k\ell}=\pi_k\pi_\ell$, for all $k\neq \ell \in U.$ The sample size is random and has a Poisson binomial distribution \citep{hod:lec:1960,ste:90,che:93}.

\subsection{Simple random sampling}

In simple random sampling (SRS) without replacement, the sample size is fixed and denoted by $n$. All the samples of size $n$ have the same probability of being selected.
The sampling design is then
$$
p(s) =
\left\{
\begin{array}{ll}
\binom{N}{n}^{-1} &\mbox{ for all }s \in \mathcal{S}_n\\
0 & \mbox{ otherwise.}
\end{array}
\right.
$$
where $\mathcal{S}_n=\{s\subset U|\#s = n \}$.
The inclusion probabilities are $\pi_k=n/N$ and $\pi_{k\ell}=n(n-1)/[N(N-1)]$, for all $k\neq \ell \in U.$

\subsection{Conditional Poisson sampling}

The problem of selecting a sample with given unequal inclusion probabilities $\pi_k$ and with fixed sample size is far from being simple. Several dozen methods have been proposed \cite[see][]{bre:han:83,til:06}. Conditional Poisson sampling (CPS) is a sampling design of fixed size and with prescribed unequal inclusion probabilities. The sampling design is given by
$$
p(s)=\frac{ \sum_{k\in S}\exp \lambda_k }{\sum_{s\in \mathcal{S}_n}\sum_{k\in S}\exp \lambda_k},
$$
where the $\lambda_k$ are obtained by solving
\begin{equation}
\sum_{s\in \{s\in\mathcal{S}_n|s\ni k\} } p(s)=\pi_k, k\in U. \label{expcps}
\end{equation}
The implementation is not simple. The complexity comes from the sum over $s\in \mathcal{S}_n$ in Expression~(\ref{expcps}) that is so large that shortcuts must be used. However, several solutions have been proposed by \citet{che:dem:liu:94} and \citet{dev:00a} in order to implement this sampling design by means of different algorithms \citep[see also][]{til:06}. The joint inclusion probabilities can easily be  computed. CPS is also called maximum entropy sampling because it maximizes the entropy as defined in Section~\ref{sec:entropy} subject to given inclusion probabilities and fixed sample size.

\subsection{Stratification}

The basic stratified sampling design consists in splitting the population into $H$ nonoverlapping strata $U_1,\dots,U_H,$ of sizes $N_1,\dots,N_H.$
Next in each stratum, a sample of size $n_h$ is selected with SRS.
The sampling design is
$$
p(s) =
\left\{
\begin{array}{ll}
\prod_{h=1}^H \binom{N_h}{n_h}^{-1} & \mbox{ for all }s\mbox{ such that } \#(U_h\cap s)=n_h,\ h=1,\dots,H,\\
0 & \mbox{ otherwise}. \\
\end{array}
\right.
$$
The inclusion probabilities are $\pi_k=n_h/N_h$ for $k \in U_h$ and
$$
\pi_{k\ell}=\left\{
\begin{array}{ll}
\frac{n_h(n_h-1)}{N_h(N_h-1)} & k,\ell\in U_h\\
\frac{n_h n_i}{N_hN_i} & k\in U_h,\ell\in U_i, i\neq h . \\
\end{array}
\right.
$$

There are two basic allocation schemes for the sample sizes:
\begin{itemize}
\item
In proportional allocation, the sample sizes in the strata are proportional to the stratum sizes in the population, which gives
$
n_h = nN_h/N.
$
Obviously $n_h$ must be rounded to an integer value.
\item
\citet{ney:34} established the optimal allocation by searching for the allocation that minimizes the variance subject to a given total sample size $n$. After some algebra, we obtain the optimal allocation:
\begin{equation}
\label{rarara}
n_h = \frac{nN_hV_h}{\sum_{\ell=1}^H N_\ell V_\ell},
\end{equation}
\end{itemize}
where
$$
V_h^2 = \frac{1}{N_h-1}\sum_{k \in U_h} \left(y_k - \overline{Y}_h\right)^2, \mbox{ and } \overline{Y}_h=\frac{1}{N_h}\sum_{k \in U_h} y_k,
$$
for $h = 1,\dots, H$. Again, $n_h$ must be rounded to an integer value. When the population is skewed, Equation~(\ref{rarara}) often gives values $n_h>N_h,$ which is almost always the case in business statistics. In this case, all the units of the corresponding stratum are selected (take-all stratum) and the optimal allocation is recomputed on the other strata. In cases where a list sampling frame is not available, the proportional and the optimal stratification might be slightly adapted.

\section{Some sampling principles}

\label{sec:principles}
The main question is how to select a sample or, in other words, what sampling method one should  use. Survey statisticians know that designing a survey is an intricate question that requires experience, a deep knowledge of the sampling frame and of the nature of variables of interest. Most sampling design manuals present a list of sampling methods. However, the choice of the sampling design should be the result of the application of several principles. In what follows, we try to establish some theoretical guidelines. Three principles can guide the choice of sample: the principle of randomization, the principle of overrepresentation and the principle of restriction.

\subsection{The principle of randomization}

\label{sec:entropy}
In design-based inference, the extrapolation of the sample estimators to the population parameters is based on the sampling design, i.e. on how the sample is selected. The first principle consists not only in selecting a sample at random but as random as possible.

A sampling design should assign a positive probability to as many samples as possible and should tend to equalize these probabilities between the samples. This enables us to avoid null joint inclusion probabilities and produces an unbiased estimator of the variance of the NHT estimator.
The common measure of randomness of a sampling design is its entropy given by
$$
I(p)=-\sum_{s\in \mathcal{S}}p(s)\log p(s),
$$
with $0\log 0=0.$

Intuitively, the entropy is a measure of the quantity of information and also a measure of randomness. High entropy sampling designs generate   highly randomized samples, which in turns make the design more robust. A discussion about high entropy designs and its relationship with robustness can be found in \citet{gra:10}.
The convergence towards asymptotic normal distributions of the estimated totals also depends on entropy. The higher the entropy is, the higher the rate of convergence is \citep{ber:98d,ber:98c}. Conversely, if the support is too small, then the distribution of the estimated total is rarely normal.

For complex sampling designs, second-order inclusion probabilities are rarely available. However, when considering high-entropy sampling designs, the variance can be estimated by using formulae that do not depend on the second-order inclusion probabilities \citep{bre:don:03}. Those estimators are approximate but are of common use. It is worth mentioning that methods for quantifying the uncertainty of complex sampling designs have been developed \citep{ant:til:11a, ber:dela:16}.

\subsection{The principle of overrepresentation}

Sampling consists in selecting a subset of the population. However, there are no particular reasons to select the units with equal inclusion probabilities. In business surveys, the establishments are generally selected with very different inclusion probabilities that are in general proportional to the number of employees.
To be efficient, the choice of units is intended to decrease uncertainty. So it is more desirable to overrepresent the units that contribute more to the variance of the estimator.

The idea of ``representativity" is thus completely misleading and is based on the false intuition that a sample must be similar to the population to perform an inference because the sample is a ``scale copy'' of the population \citep[see among others][]{kru:mos:79a,kru:mos:79b,kru:mos:79c}. In fact, the only requirement for the estimator to be unbiased consists of using a sampling design with non-null first-order inclusion probability for all units of the population, which means that the sampling design does not have coverage problems \citep[see][p.~8]{sar:swe:wre:92}.
Unequal probability sampling can be used to estimate the total $Y$ more efficiently. The main idea is to oversample the units that are more uncertain because the sample must collect as much information as possible from the population, which was already the basic idea of the seminal papers of Jerzy \citet{ney:34,ney:38} on optimal stratification. In general, the principle of overrepresentation implies that a sampling design should have unequal inclusion probabilities if prior information is available. There exist different ways to deduce the inclusion probabilities from a working model as we will see in Section \ref{sec:model-assis}. 

\subsection{The principle of restriction}
The principle of restriction consists in selecting only samples with a given set of characteristics, for instance, by fixing the sample size or the sample sizes in categories of the population (stratification).  There are many reasons why restrictions should be imposed. For instance, empty categories in the sample might be avoided, which can be very troublesome when the aim is to estimate parameters in small subsets of the population. It is also desirable that the estimates from the sample are coherent with some auxiliary knowledge. So only samples that satisfy such a property can be considered. By coherent, we mean that the estimate from the sample of an auxiliary variable should match a known total. Such samples are said to be balanced. Balanced sampling is discussed Section~\ref{sec::balance}. More generally, restrictions can reduce or even completely remove the dispersion of some estimators.

At first glance, the principle of restriction seems to be in contradiction with the principle of randomization because it restricts the number of samples with non-null probabilities. However the possible number of samples is so large that, even with several constraints, the number of possible samples with non-null probabilities can remain very large. It is thus still reasonable to assume a normal distribution for the estimates. Balanced sampling enables us to avoid the ``bad'' samples, which are those that give estimates for the auxiliary variables that are far from the known population totals.

\section{Balanced sampling}
\label{sec::balance}
A sample without replacement from a population of size $N$ can be denoted by a vector of size $N$ such that the $k$th component is equal to $1$ if the $k$th unit is selected and $0$ if it is not. Following this representation, a sample can be interpreted as a vertex of the unit hypercube of dimension $N$. This geometrical interpretation of a sample is central in the development of some sampling algorithms \citep{til:06}.

By definition, a balanced sample satisfies
\begin{equation}
\sum_{k\in S} \frac{\xb_k}{\pi_k} =\sum_{k\in U} \xb_k.
\label{eq:balanced}
\end{equation}
where $\xb_k = (x_{k1}, \dots, x_{kp})^\top$ is a vector of $p$ auxiliary random variables measured on unit $k$.
Vectors $\xb_k$ are assumed to be known for each unit of the population, i.e. a register of population is available for the auxiliary information.
The choice of the first-order inclusion probabilities is discussed in Section \ref{sec:model-assis} and is a consequence of the principle of overrepresentation. Balanced sampling designs are designs whose support is restricted to samples satisfying (or approximately satisfying) Equation~(\ref{eq:balanced}). In other words, we are considering sampling designs of prescribed first-order inclusion probabilities $\pi_1,\dots, \pi_N$ and with support
$$ \left\{s\in \mathcal{S} : \sum_{k\in s} \frac{\xb_k}{\pi_k} =\sum_{k\in U} \xb_k \right\}.$$
More generally, an approximately balanced sample $s$ is a sample satisfying
\begin{equation}
\label{eq:app_balanced}
\left\|\mathbf{D}^{-1}\left( \sum_{k\in U} \xb_k -\sum_{k\in s} \xb_k/\pi_k\right)\right\| \leq c,
\end{equation}
where $\mathbf{D}$ is a $p\times p$ matrix defined by $\mathbf{D}=\text{diag}(\sum_{k\in U}  \xb_k)$, $c$ is a positive constant playing the role of a tolerance from the deviation of the balancing constraints and $\|\cdot\|$ denotes any norm on $\mathbb{R}^p$.
Balanced sampling thus consists in selecting randomly a sample whose NHT-estimators are equal or approximately equal to the population totals for a set of auxiliary variables. In practice, exact balanced sampling designs rarely exist. The advantage of balanced sampling is that the design variance is null or almost null for the NHT-estimators for these auxiliary variables. Thus, if the variable of interest is strongly correlated with these auxiliary variables, the variance of the NHT-estimator of the total for the variable of interest is also strongly reduced.

Sampling designs with fixed sample size (SRS, CPS) and stratification are particular cases of balanced sampling. Indeed in sampling with fixed sample size the only balancing variable is the first-order inclusion probability $\xb_k=\pi_k$ for all $k\in U.$ In stratification, the $H$ balancing variables are $\xb_k=(\pi_kI(k\in U_1),\dots,\pi_kI(k\in U_h),\dots,\pi_kI(k\in U_H))^\top$, where $I(k\in U_h)$ is the indicator variable of the presence of unit $k$ in stratum $U_h.$

The first way to select a balanced sample could consist in using a rejective procedure, for instance by  generating samples with SRS or Poisson sampling until a sample satisfying Constraint~(\ref{eq:app_balanced}) is drawn. However, a conditional design does not have the same inclusion probabilities as the original one. For instance, \citet{leg:yu:10} have shown that a rejective procedure fosters the selection of central units. So the units with extreme values have smaller inclusion probabilities. Well before, \citet{haj:81} already noticed that if samples with a Poisson design are generated until a fixed sample size is obtained, then the inclusion probabilities are changed. This problem was solved by \citet{che:dem:liu:94} who described the link between the Poisson design and the one obtained by conditioning on the fixed sample size \citep[see also][pp. 79-96]{til:06}. Unfortunately, the computation of conditional designs seems to be intractable when the constraint is more complex than fixed sample size. Thus the use of rejective methods cannot lead to a sampling design whose inclusion probabilities are really computable.

The cube method \citep{dev:til:04a} allows us to select balanced samples at random while preserving the possibly unequal prescribed first-order inclusion probabilities. The method starts with the prescribed vector of inclusion probabilities. This vector is then randomly modified at each step in such a way that at least one component is changed to 0 or 1 and such that this transformation respects the prescribed first-order inclusion probabilities. Thus the cube algorithm sequentially selects  a sample in at most $N$ steps. At each step, the random modification is realized while respecting the balancing constraints and the inclusion probabilities. The algorithm has two distinct phases: the first is the flight phase, where the balanced equations are exactly satisfied. At some point, it is possible that the balancing equations can only be approximated. In the second phase, called landing phase, the algorithm selects a sample that nearly preserves the balancing equation while still exactly satisfying the prescribed inclusion probabilities.

It is not possible to fully characterize the sampling design generated by the cube method. In particular, second-order inclusion probabilities are intractable. In order to compute the variance, \citet{dev:til:05} gave several approximations using only first-order inclusion probabilities. \citet{bre:cha:11} suggest using a martingale difference approximation of the values of $\Delta_{k\ell}$ that takes into account the variability of both the flight and the landing phase, unlike the estimators proposed by \citet{dev:til:05}.

The cube method has been extended by \citet{til:fav:04} to enable the coordination of balanced samples. Optimal inclusion probabilities are studied in the perspective of balanced sampling by \citet{til:fav:05} and are further investigated by \citet{cha:bon:dev:opt:11}.
\citet[][in French]{dev:14} sketches a proof of the conditions that must be met on the inclusion probabilities and on the auxiliary variables in order to achieve an exact balanced sample. In this case, the cube algorithm only has a flight phase. 
From a practical standpoint, several implementations exist in the R language \citep{til:mat:15,graf2016} and in SAS \citep{rou:tar:04,cha:til:05b}.

\section{Model-assisted choice of the sampling design and balanced sampling}
\label{sec:model-assis}

\subsection{Modelling the population}
The  principles of overrepresentation and restriction can be implemented through a modelling of the links between the variable of interest and the auxiliary variables. This model may be relatively simple,for instance a linear model:
\begin{equation}
y_k = \xb_k^\top \betag+\varepsilon_k,\label{M1}
\end{equation}
where $\xb_k=(x_{k1},\dots,\dots,x_{kp})^\top$ is a vector of $p$ auxiliary variables, $\betag$ is a vector of regression coefficients, and $\varepsilon_k$
are independent random variables with null expectation and variance $\sigma_{\varepsilon k}^2.$ The model thus admits heteroscedasticity. The error terms $\varepsilon_k$ are supposed to be independent from the random sample $S$. Let also $\E_M(.)$ and $\Var_M(.)$ be respectively the expectation and variance under the model.

Under model~(\ref{M1}), the anticipated variance of the NHT-estimator is
$$
{\rm AVar}(\widehat{Y})=\E_p\E_M(\widehat{Y}-Y)^2=
\E_p\left( \sum_{k\in S}\frac{\xb_k^\top \betag}{\pi_k}-\sum_{k\in U} \xb_k^\top \betag \right)^2 + \sum_{k\in U}  (1-\pi_k)\frac{\sigma^2_k}{\pi_k}.
$$
The second term of this expression is called the Godambe-Joshi bound \citep{god:jos:65}.

Considering the anticipated variance, for a fixed sample size $n$, the sampling design that minimizes the anticipated variance consists in
\begin{itemize}
\item using inclusion probabilities proportional to $\sigma_{\varepsilon k}$,
\item using a balanced sampling design on the auxiliary variables $\xb_k$.
\end{itemize}
The inclusion probabilities are computed using
\begin{equation}
\pi_k=\frac{n \sigma_{\varepsilon k}}{\sum_{\ell\in U} \sigma_{\varepsilon \ell}},
\label{calacul}
\end{equation}
provided that $
n \sigma_{\varepsilon k} < \sum_{\ell\in U} \sigma_{\varepsilon \ell}$ for all $k\in U.$ If it is not the case, the corresponding inclusion probabilities are set to one and the inclusion probabilities are recomputed according to Expression~(\ref{calacul}).

If the inclusion probabilities are proportional to $\sigma_{\varepsilon k}$ and the sample is balanced on the auxiliary variables $\xb_k$, the anticipated variance becomes \citep{ned:til:08}
$$
N^2 \left[ \frac{N-n}{N}\frac{\bar{\sigma}_\varepsilon^2}{n} - \frac{\Var(\sigma_{\varepsilon})}{N}\right],
$$
where
$$
\bar{\sigma}_\varepsilon = \frac{1}{N} \sum_{k\in U} \sigma_{\varepsilon k}
\mbox{ and }\Var(\sigma_{\varepsilon}) = \frac{1}{N} \sum_{k\in U}\left( \sigma_{\varepsilon k}- \bar{\sigma}_\varepsilon \right)^2.
$$
Applying the randomization principle would result in a maximum entropy sampling design under the constraint of minimizing the anticipated variance. However, except for the very particular cases given in Table~\ref{tab1}, there is no known general solution to this problem. When it exists, we refer to such a sampling design as ``optimal''. All the designs presented in Section~\ref{sec:basic} are optimal for particular cases of Model~(\ref{M1}) and are explicitly described in Table~\ref{tab1}.

\begin{table}[htb!]
\begin{center}
\caption{Particular cases of Model~(\ref{M1}) and corresponding optimal sampling design\label{tab1}}
\begin{tabular}{llcl}
\hline
Underlying Model & Design & Model Variance & $\pi_k$ \\
\hline
$y_k=\beta+\varepsilon_k$ & SRS & $\sigma^2$ & $n/N$ \\
$y_k= \varepsilon_k $ & Bernoulli sampling & $\sigma^2$ & $\pi=\E(n_S)/N$ \\
$y_k=x_k\beta+\varepsilon_k$ & CPS & $ x_k^2 \sigma^2 $ & $\pi_k\propto x_k$ \\
$y_k= \varepsilon_k $ & Poisson sampling & $x_k^2 \sigma^2$ & $\pi_k\propto x_k$ \\
$y_k= \beta_h+ \varepsilon_k,k\in U_h, $ & Proportional stratification & $\sigma^2$ & $n/N$ \\
$y_k= \beta_h+ \varepsilon_k ,k\in U_h, $ & Optimal stratification & $\sigma_h^2$ & $\pi_k\propto \sigma_h $\\
\hline
\end{tabular}
\end{center}
\end{table}

Maximizing entropy tends to equalize the probabilities of selecting samples. For Bernoulli sampling all the samples of the same size have the same probability of being selected. Under SRS, all the samples with a non-null probability have exactly the same probability of being selected.
In stratification, the support is reduced to the samples with fixed sample sizes in each stratum. Curiously, all the samples of the support have the same probability of being selected even with optimal stratification.

When all the inclusion probabilities are unequal, it is not possible to equalize the probabilities of the samples. However, in CPS, all the samples of size $n$ have positive probabilities of being drawn and, in Poisson sampling, all the samples (of any size) have non-null probabilities. Even though the inclusion probabilities are not equal, all the samples in the support may have the same probability of being selected, for instance in optimal stratification.

The common sampling designs presented in Table~\ref{tab1} correspond to very simple models. For SRS, the model only assumes a parameter $\beta$ and homoscedasticity. In stratification, the means $\beta_h$ of the variable of interest can be different in each stratum. Moreover for optimal stratification, the variances $\sigma_h^2$ of the noise $\varepsilon_k$ are presumed to be different in each stratum. Unfortunately, there is no general algorithm that enables us to implement an unequal probability balanced sampling design with maximum entropy for the general case of Model~(\ref{M1}).

The cube method is still not a fully complete optimal design for the general Model~(\ref{M1}) because the entropy is not maximized. The cube method however gives a solution to a general problem, which involves SRS, unequal probability sampling with fixed sample size and stratification that are all particular cases of balanced sampling. Even if it is not possible to maximize the entropy with the cube method, it is possible to randomize the procedure, for instance, by randomly sorting the units before applying the algorithm.

\subsection{A link with the model-based approach}
An alternative literature is dedicated to the model-based approach. In this framework, the problem of estimating a total is seen as a prediction problem and the population is modelled. The model-based approach assumes a super-population model and the inference is carried out using this model \citep{bre:63a,roy:70b,roy:70,roy:76b,roy:76,roy:92, roy:her:73a, roy:her:73b}. The Best Linear Unbiased Predictor (BLUP) under Model~(\ref{M1})
is given by
$$
\widehat{Y}_{BLU} = \sum_{k\in S} y_k + \sum_{k\in U\backslash S} \xb_k^\top \widehat{\betag},
$$
where
$$
\widehat{\betag} =
\left(\sum_{k\in S} \frac{\xb_k^\top \xb_k}{\sigma_{\varepsilon k}^2} \right)^{-1} \sum_{k\in S}
\frac{\xb_k^\top y_k}{\sigma_{\varepsilon k}^2},
$$
\citep[see also][]{val:dor:roy:00}.

\citet{ned:til:08} show that, under the assumption that there are two vectors $\lambdag$ and $\gammag$ of $\R^p$ such that $\lambdag^\top \xb_k=\sigma_{\varepsilon k}^2$ and $\gammag^\top \xb_k=\sigma_{\varepsilon k}$ for all $k\in U$, then, for the sampling design that minimizes the anticipated variance, the NHT-estimator is equal to the BLUP. In this case, both approaches coincide. So, in this respect, balanced sampling enables us to reconcile design-based and model-based approaches.

\subsection{Beyond the linear regression model}
A generalization of Model~(\ref{M1}) is the linear mixed model \citep[see among others][]{jiang:07,rup:wand:car:03}. It encompasses many widely used models and is extensively used in survey sampling, especially in the field of small area estimation. \citet{brei:chauv:12} investigate the application of balanced sampling in the case where the working model is a linear mixed model and they introduce penalized balanced sampling in order to into account the random effects.
Linear-mixed as well as nonparametric model-assisted approaches have been extensively studied in the context of survey sampling \citep{brei:ops:2016}.

\section{Spatial balanced sampling}
\label{sec::spat-samp}
\subsection{Modeling the spatial correlation}

Spatial sampling is particularly important in environmental statistics. A large number of specific methods were developed for environmental and ecological statistics \citep[see among others][]{marker2009sampling,tho:12}.

When two units are geographically close, they are in general similar, which induces a spatial dependency between the units.
Consider the alternative model
\begin{equation}
\label{M2}
y_k = \xb_k^\top \betag+\varepsilon_k,
\end{equation}
where $\xb_k=(x_{k1},\dots,\dots,x_{kp})^\top$ is a set of $p$ auxiliary variables, $\betag$ is a vector of regression coefficients, and $\varepsilon_k$ are random variables with $\E(\varepsilon_k)=0,$ $\Var(\varepsilon_k)=\sigma^2_k$ and $\Cov(\varepsilon_k,\varepsilon_\ell)=\sigma_{\varepsilon k}\sigma_{\varepsilon_\ell} \rho_{k\ell}$. The model admits heteroscedasticity and autocorrelation.
   The error terms $\varepsilon_k$ are supposed to be independent from the random sample $S$. Let also $\E_M(.)$ and $\Var_M(.)$ be respectively the expectation and variance under the model.

Under model (\ref{M2}), \citet{gra:til:13} show that the anticipated variance of the NHT-estimator is
\begin{equation}
\label{eq:ant_var_spat}
{\rm AVar}(\widehat{Y})=
\E_p\left( \sum_{k\in S}\frac{\xb_k^\top \betag}{\pi_k}-\sum_{k\in U} \xb_k^\top \betag \right)^2 + \sum_{k\in U}\sum_{\ell\in U}\Delta_{k\ell}\frac{\sigma_{\varepsilon k} \sigma_{\varepsilon \ell} \rho_{k\ell}}{\pi_k \pi_\ell}.
\end{equation}
If the correlation $\rho_{k\ell}$ is large when the units are close, the sampling design that minimizes the anticipated variance consists in
\begin{itemize}
\item using inclusion probabilities proportional to $\sigma_{\varepsilon k}$,
\item using a balanced sampling design on the auxiliary variables $\xb_k$,
\item and avoiding the selection of neighboring units, i.e. selecting a well spread sample (or spatially balanced).
\end{itemize}

If the selection of two neighboring units is avoided, the values of $\Delta_{k\ell}$ can be highly negative, which makes the anticipated variance~(\ref{eq:ant_var_spat}) small.

The value of $\Delta_{k\ell}$ can be interpreted as an indicator of the spatial pairwise behavior of the sampling design. Indeed, if two units $k$ and $\ell$ are chosen independently with inclusion probability $\pi_k$ and $\pi_\ell$ respectively, then the joint inclusion probability is $\pi_k \pi_\ell$. Hence, if $\Delta_{k\ell} <0$, respectively $\Delta_{k\ell} > 0$, the sampling design exhibits some repulsion, respectively clustering, between the units $k$ and $\ell$. In other words, the sign of $\Delta_{k\ell}$ is a measure of repulsion or clustering of the sampling design for two units $k$ and $\ell$. Similar ideas have been used in the literature on spatial point processes to quantify the repulsion of point patterns. In particular, the pair correlation function is a common measure of the pairwise interaction \citep[][chap. 4]{Moller03}.

\subsection{Generalized Random Tessellation Stratified Design}
The {\it Generalized Random Tessellation Stratified} (GRTS) design was proposed by \citet{stevens1999spatially,Stev:Olse:spat:2004,Stev:Olse:vari:2003}. The method is based on the recursive construction of a grid on the space. The cells of the grid must be small enough so that the sum of the inclusion probabilities in a square is less than 1. The cells are then randomly ordered such that the proximity relationships are preserved. Next a systematic sampling is applied along the ordered cells. The method is implemented in the ``spsurvey'' R package \citep{sp}.

\subsection{Local pivotal method\label{spatialpivotal}}
The pivotal method has been proposed by \citet{dev:til:00} and consists in selecting two units, say $i$ and $j$, with inclusion probabilities $0< \pi_i, \pi_j <1$ in the population at each step and randomly updating their inclusion probability in order to set $\pi_i$ or $\pi_j$ to $0$ or $1$, while preserving in expectation the original inclusion probabilities.
If the two units are sequentially selected according to their order in the population, the method is called \emph{sequential pivotal method} (or \emph{ordered pivotal sampling} or \emph{Deville's systematic sampling}) \citep{cha:12}. The sequential pivotal method is also closely related to the sampling designs introduced by \citet{ful:70}.

\citet{gra:lun:sch:12} have proposed using the pivotal method for spatial sampling. This method is called {\it local pivotal sampling} and the two competing units are neighbors. If the probability of one of these two units is increased, the probability of the other is decreased, which in turn induces some repulsion between the units
and the resulting sample is thus well spread.

\subsection{Spreading and balancing: local cube method}
In the local pivotal, two units compete to be selected. The natural extension of this idea is to let a cluster of units fight. The local pivotal method has been generalized by \citet{gra:til:13} to provide a sample that is at the same time well spread in space and balanced on auxiliary variables in the sense of Expression~(\ref{eq:balanced}). This method, called {\it local cube}, consists in running the flight phase of the cube method on a subset of $p+1$ neighboring units, where $p$ is the number of auxiliary variables. After this step, the inclusion probabilities are updated such that:
\begin{itemize}
\item one of the $p+1$ units has its inclusion probability updated to $0$ or $1$,
\item the balancing equation is satisfied.
\end{itemize}
When a unit is selected, it decreases the inclusion of the $p$ other units of the cluster. Hence, it induces a negative correlation in the selection of neighboring units, which in turn spreads the sample.

\subsection{Spatial sampling for non-spatial problems}

Spatial methods can also be used in a non-spatial context. Indeed, assume that a vector ${\bf x}_k$ of auxiliary variables is available for each unit of the population. These variables can be, for instance, turnover, profit or the number of employees in business surveys. Even if these variables are not spatial coordinates, they can be used to compute a distance between the units. For instance, the Mahalanobis distance can be used:
$$
d^2(k,\ell) =(\xb_k-\xb_\ell)^\top \Sigmag^{-1} (\xb_k-\xb_\ell),
$$
where
$$
\Sigmag =\frac{1}{N} \sum_{k\in U} (\xb_k-\bar{\xb})(\xb_k - \bar{\xb})^\top
\mbox{ and }
\bar{\xb} =\frac{1}{N} \sum_{k\in U} \xb_k .
$$
\citet{gra:lun:sch:12} advocate the use of spreading on the space of the auxiliary variables. Indeed, if the response variable is correlated with the auxiliary variable, then spreading the sample on the space of auxiliary variables also spreads the sampled response variable. It also induces an effect of smooth stratification on any convex set of the space of variables. The sample is thus stratified for any domain, which can be interpreted as a property of robustness.

\section{Illustrative example}
\label{sec:ill-examp}

\begin{figure}[htb!]
\begin{center}
\includegraphics[scale=0.30]{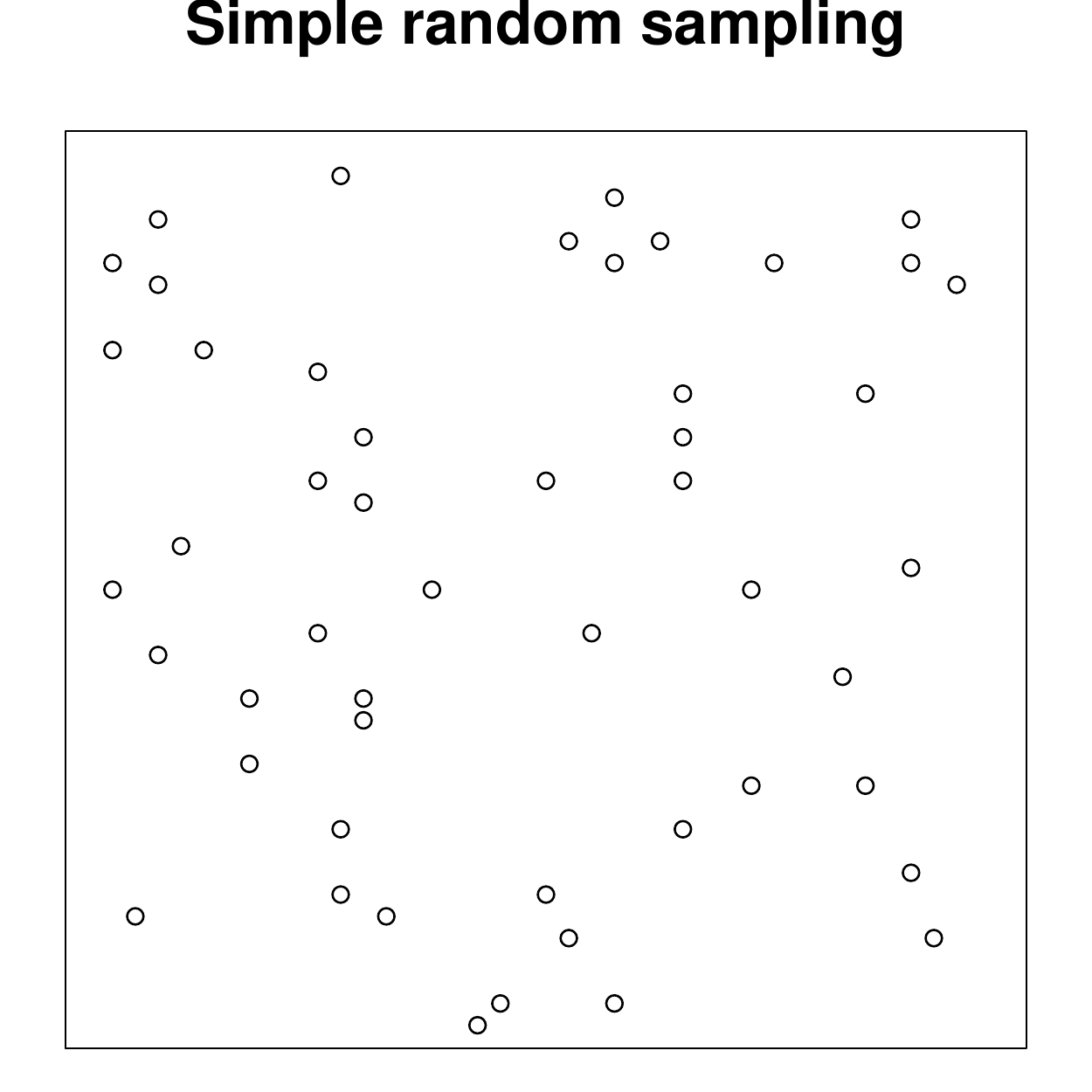}
\includegraphics[scale=0.30]{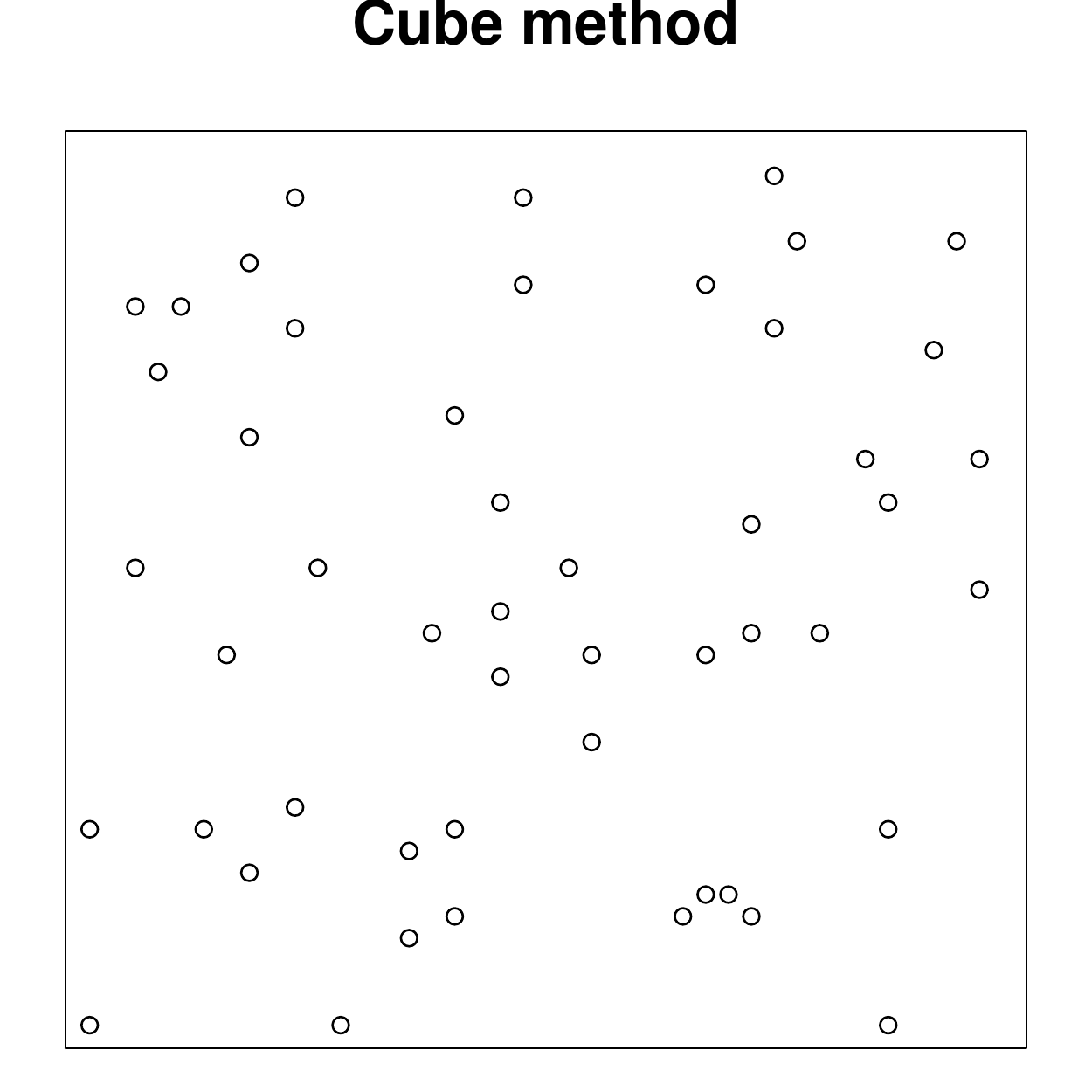}
\end{center}
\caption{Sample of size $n=50$ in a population of 1600 plots by means of SRS and the cube method. These samples are not very well spread.\label{t1}}
\end{figure}

A simple example illustrates the advantages of the sampling designs discussed in Section~\ref{sec::spat-samp}. Consider a square of $N=40\times 40 = 1600 $ dots that are the sampling units of the population.
A sample of size $n=50$ is selected from this population by means of different sampling designs. Figure~\ref{t1} contains two samples that are not spatially balanced: SRS and balanced sampling by means of the cube method. For balanced sampling, three variables are used: a constant equal to 1, the $x$ coordinate and the $y$ coordinate. So the sample has a fixed sample size and is balanced on the coordinates. These samples are not well spread or spatially balanced.

\begin{figure}[htb!]
\begin{center}
\includegraphics[scale=0.30]{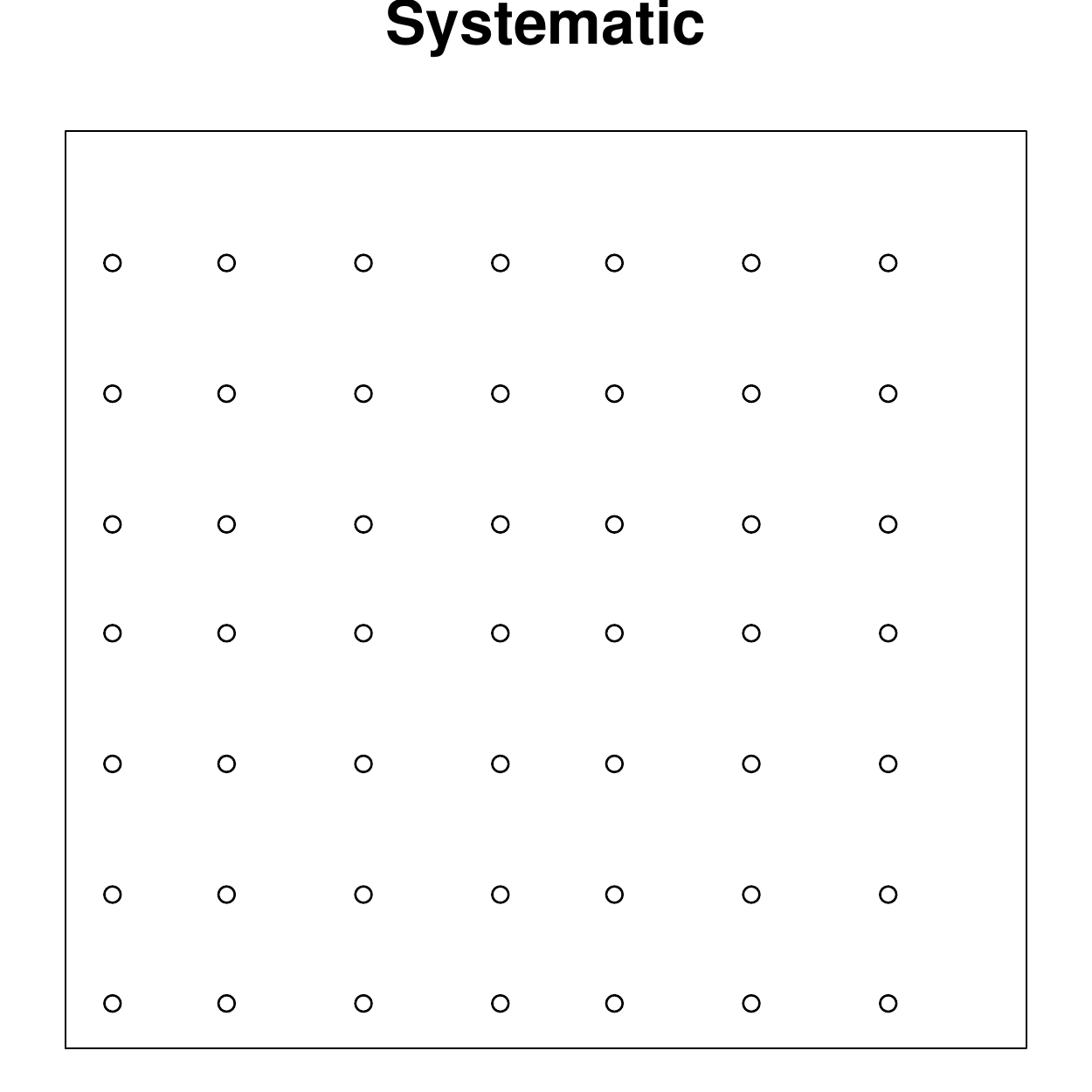}
\includegraphics[scale=0.30]{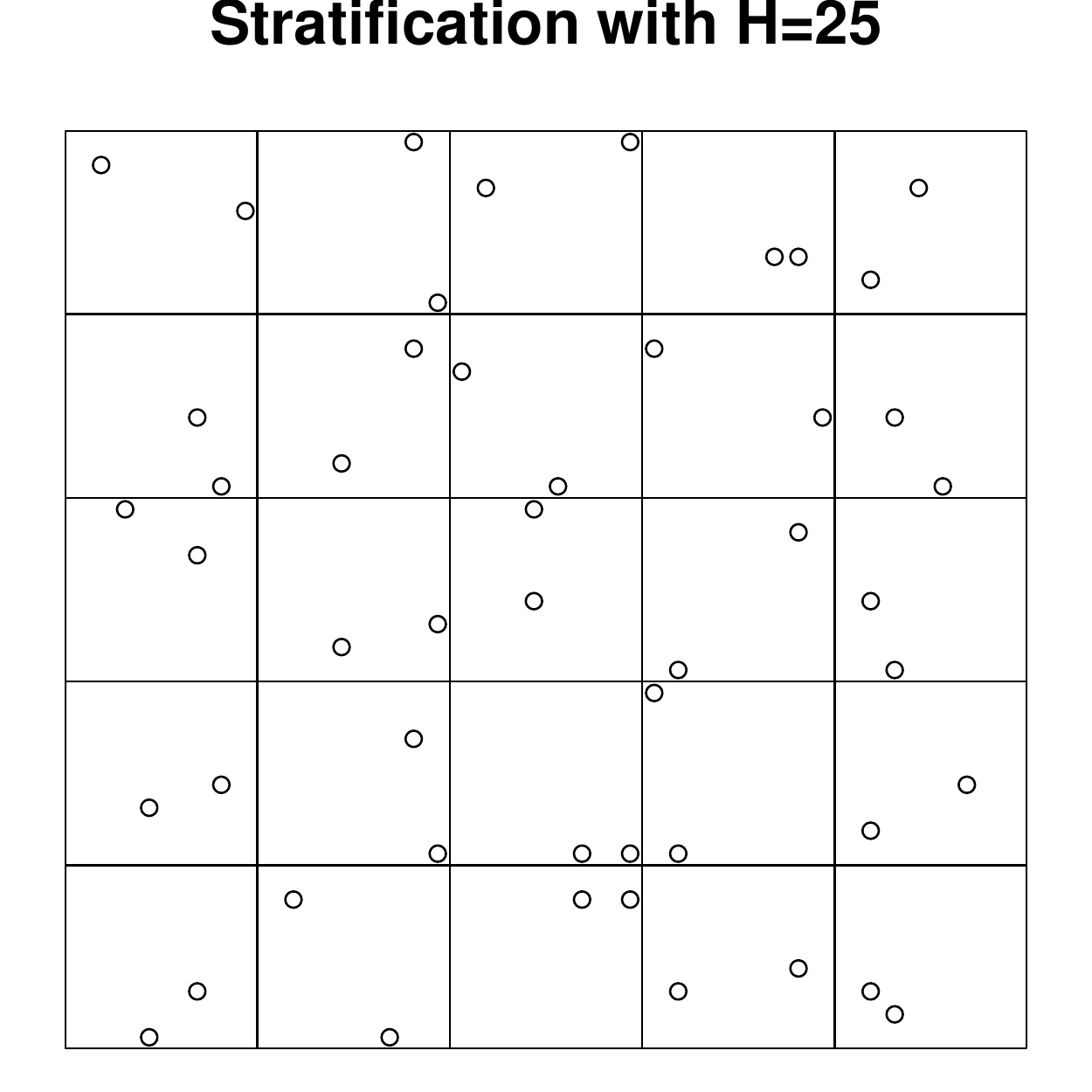}
\end{center}
\caption{Systematic sampling and stratification with 2 units per stratum.\label{t2}}
\end{figure}

Figure~\ref{t2} contains the most basic sampling designs used to spread a sample: systematic sampling and stratification.
Unfortunately, spatial systematic sampling cannot be generalized to unequal probability sampling. It is also not possible to apply it to a population on a lattice.

\begin{figure}[htb!]
\begin{center}
\includegraphics[width=0.32\textwidth]{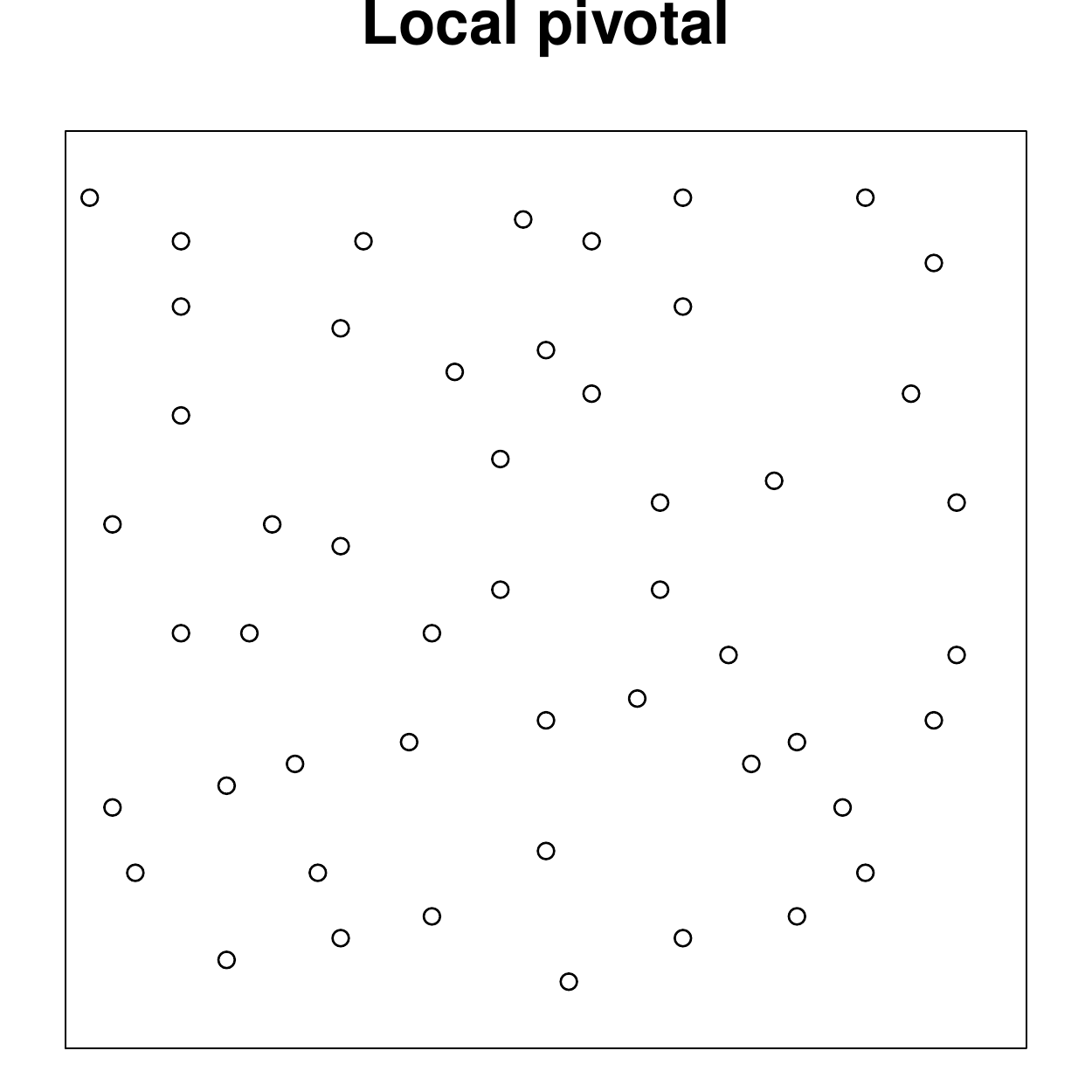}
\includegraphics[width=0.32\textwidth]{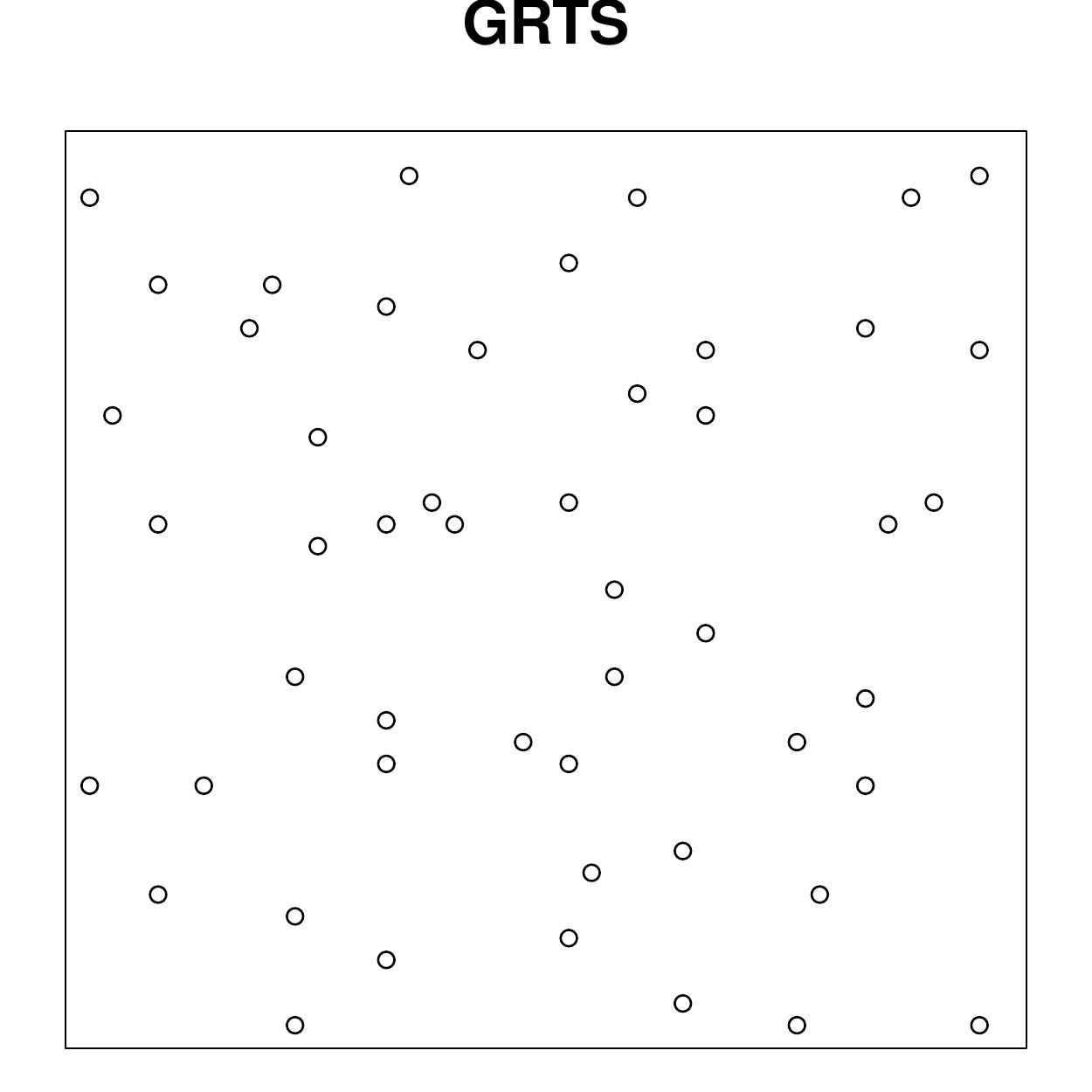}
\includegraphics[width=0.32\textwidth]{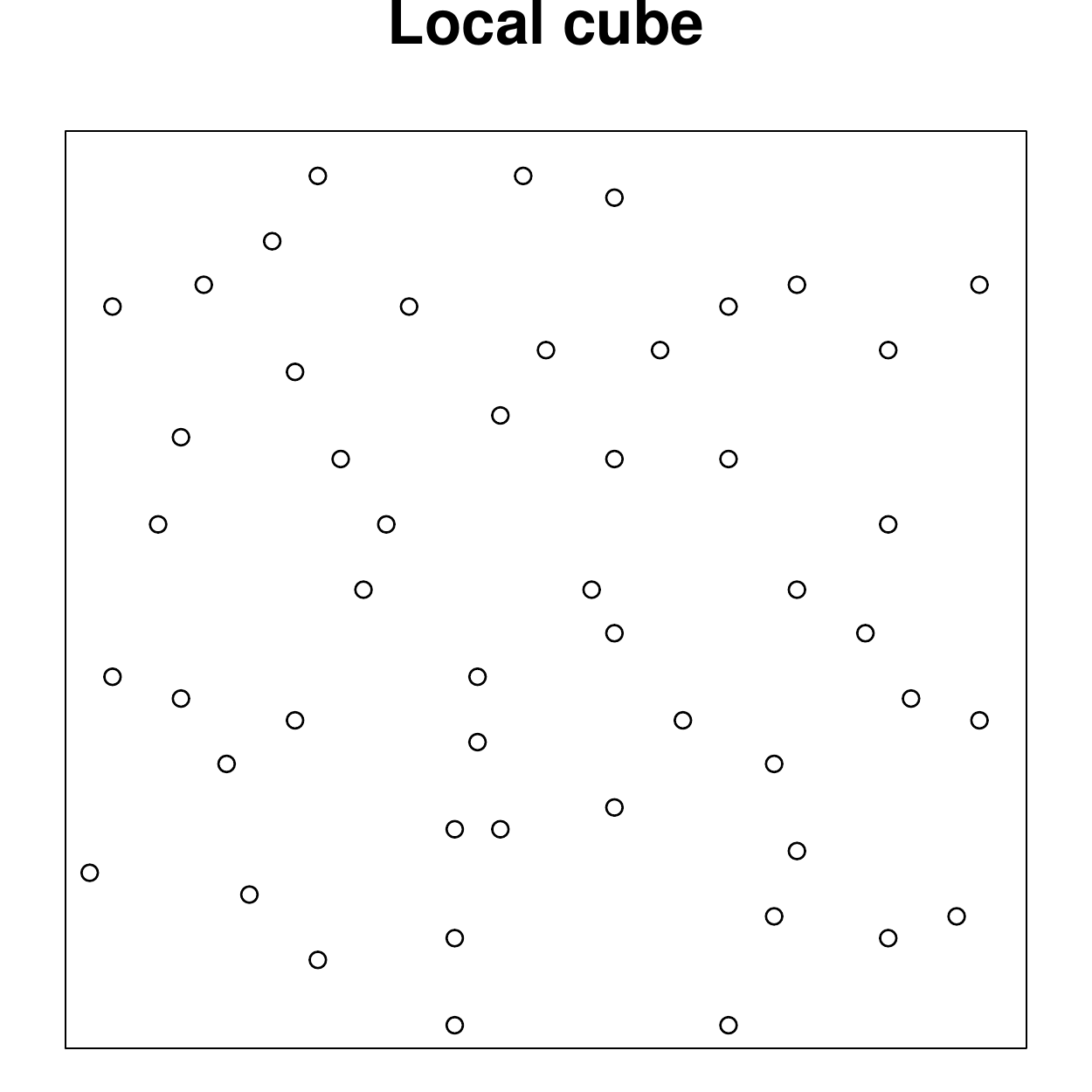}
\end{center}
\caption{GRTS, pivotal method, and local cube method. Samples are well spread.\label{t3}}
\end{figure}

Figure~\ref{t3} contains modern well spread sampling methods such as the local pivotal method, GRTS and the local cube method. At first glance, it is difficult to evaluate which design gives the most well spread sample.

\begin{figure}[htb!]
\begin{center}
\includegraphics[width=0.32\textwidth]{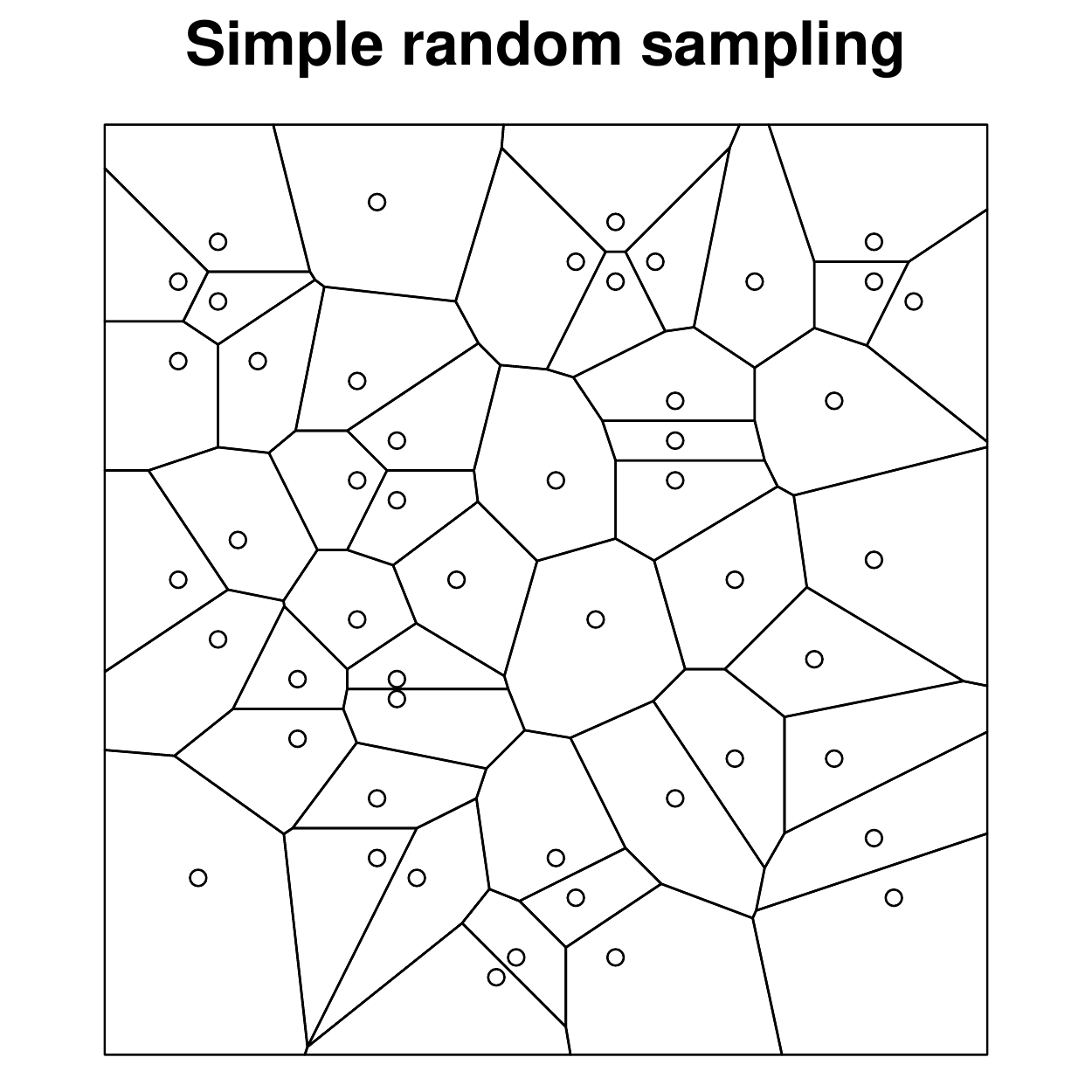}
\includegraphics[width=0.32\textwidth]{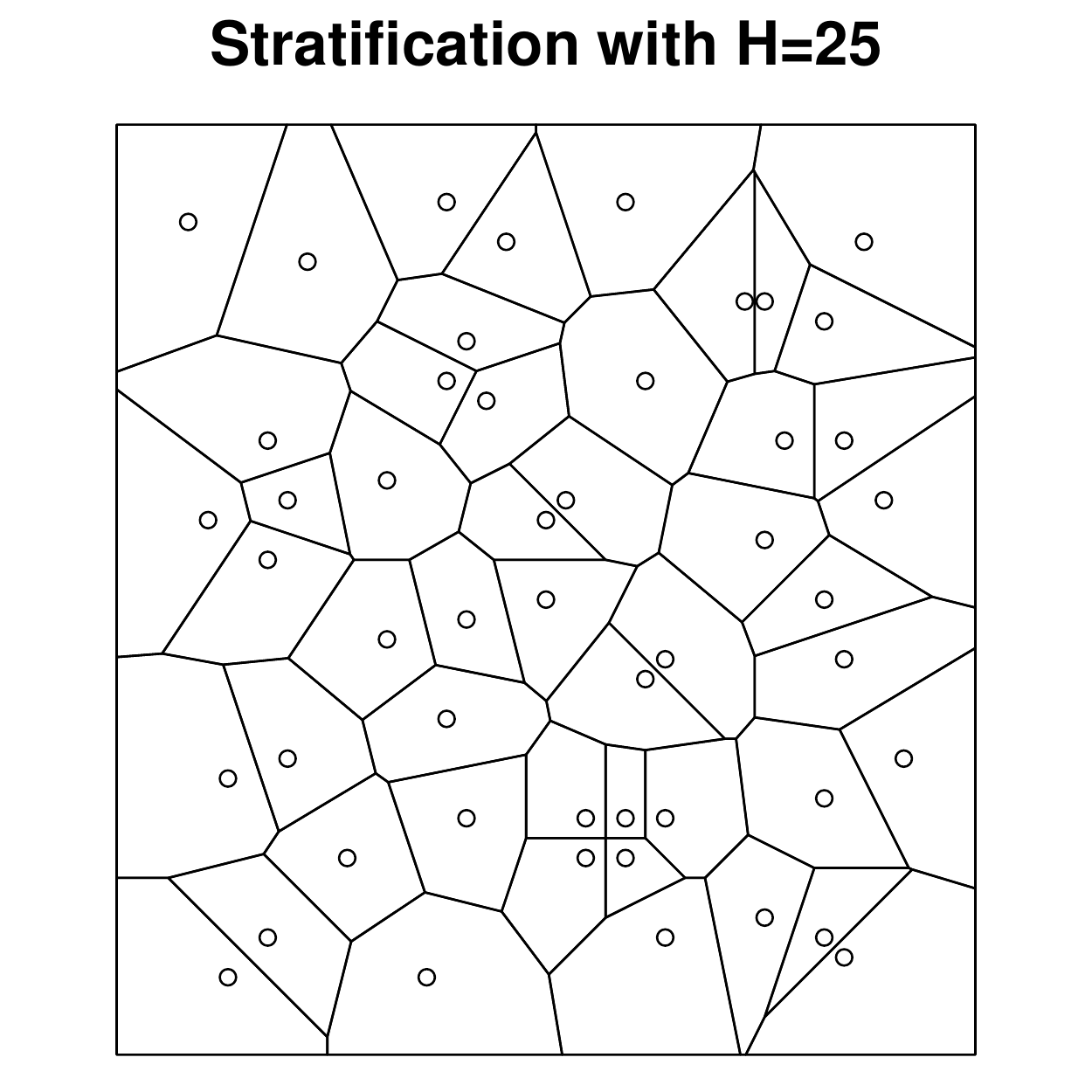}
\includegraphics[width=0.32\textwidth]{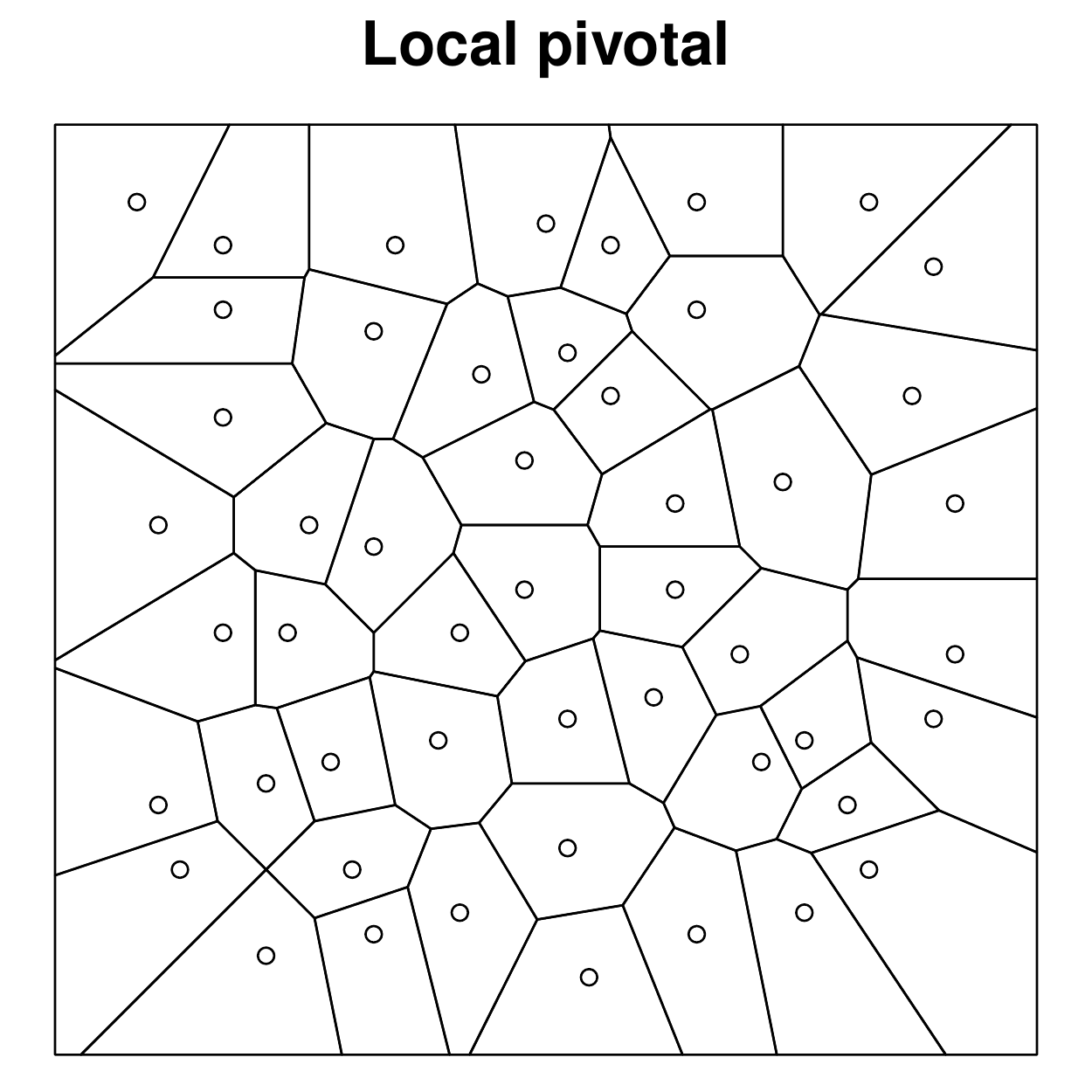}
\end{center}
\caption{Example of Vorono{\"\i} polygons for three sampling designs.\label{t4}}
\end{figure}

A Vorono{\"\i} polygon is the set of elements of the population that are closer to a given point than any other points in the population. Figure~\ref{t4} contains the Vorono{\"\i} polygons for SRS, stratification and local pivotal method.
 The variance of the sum of inclusion probabilities of the population units that are included in a polygon is an indicator of the quality of spatial balancing \citep{Stev:Olse:spat:2004}.

Table~\ref{ttt1} contains the average of the indicators of spatial balance for 1000 selected \citep{graf:lund:13}.
For systematic sampling, the index is not null because of the edge effect. The best designs are the local pivotal methods and the local cube method.
\begin{table}[ht]
\centering
\caption{Indices of spatial balance for the main sampling designs\label{ttt1} with their standard deviations}
\begin{tabular}{lr}
\hline
Design & Balance indicator \\
\hline
Systematic & 0.05 (0.04) \\
Simple random sampling & 0.31 (0.10) \\
Stratification with H=25 & 0.10 (0.02) \\
Local pivotal & 0.06 (0.01) \\
Cube method & 0.21 (0.06) \\
Local Cube method & 0.06 (0.01) \\
GRTS & 0.10 (0.02) \\
\hline
\end{tabular}
\end{table}
\citet{graf:lund:13} extensively discuss the concept of balancing and the implication on the estimation. In particular, they show under some assumptions that a well spread sampling design is an appropriate design under model~(\ref{M2}).
\section{Discussion}
\label{sec:disc}
Three principles theoretically appealing have been established. Modelling the population can be used as a tool for the implementation of the principle of overrepresentation and of restriction. Indeed, the use of auxiliary variables through a model determines the inclusion probabilities (overrepresentation) and imposes a balancing condition (restriction). Thus, balanced sampling is a crucial tool to implement these principles.

However, some limitations of the scope of this paper must be outlined. First, it is worth noting that beyond the theoretical principles there are also a practical constraints. Practitioners have to take the context into account. A very large number of practical issues affect the direct applications of the suggested theoretical principles. So we recommend to keep those principles in mind when designing a survey even though we acknowledge that it is probably not always possible to apply them because of constraints such as time, inaccurate sampling frame or budget.

In addition to this, a simplicity principle can be predominant. A large number of environmental monitoring surveys are based on a systematic spatial sampling just because this design has the advantage of being simple, spread and easy to implement.

Moreover, in the case of multi-objective surveys, a single model that summarizes the link between the variables of interest and the auxiliary variables is not always available. There is sometimes an interest for regional or local estimations or for complex statistics. The aim can thus not be reduced to the estimation of a simple total. Compromises should then be established between the different objectives of the samples \citep{falorsi2008balanced,falorsi2016unified}.

Finally, surveys are also repeated in time, which makes the problem much more intricate. Cross-sectional and longitudinal estimations require very different sampling designs. It is always better to select the same sample to estimate evolutions, while for transversal estimations independent samples are more efficient. In business statistics, great attention is given to the survey burden that should be fairly distributed between  companies. For these reasons, in a large number of surveys, statisticians foster a partial rotation of the units in the sample. Rotation is sometimes difficult to reconcile with the optimization of the transversal designs.

The three principles formalized and developed in this paper should guide the choice of the sampling design whenever possible. The principle of randomization should always be considered by trying to maximize the entropy, possibly under some constraints. The other two principles can only be applied when the population is explicitly modelled. This modelling may or may not be used as an assumption for the inference, depending on whether a design-based or a model-based approach is adopted.
Using a model-assisted approach, we advocate the use of a model to apply the principles of overrepresentation and of restriction while preserving the design unbiasedness of the Horvitz-Thompson estimator.

\section*{Acknowledgements}
The authors would like to thank two referees and the Associate Editor for constructive remarks that have helped to greatly improve the paper. We would like to thank Mihaela Anastasiade for a careful reading and Audrey-Anne Vall\'ee for useful comments on an early draft of the present paper.

\end{document}